\documentclass[twocolumn,showkeys,showpacs,preprintnumbers,prd,superscriptaddress,nofootinbib]{revtex4}
\usepackage{graphicx}
\usepackage{epsf}
\usepackage{bm}
\usepackage{amsmath}
\usepackage{amsfonts}
\usepackage{amssymb}
\usepackage{epstopdf}
\usepackage{natbib}
\usepackage{color}
\usepackage{verbatim}
\usepackage{multirow}
\usepackage{bm}
\usepackage{listings}
\usepackage{soul}
\usepackage{caption}
\usepackage{subcaption}
\usepackage{ragged2e}
\usepackage{float}
\usepackage{hyperref}

\begin{document}

\title{Non-parametric reconstruction of the fine structure constant with galaxy clusters}

\author{Marcelo Ferreira}
\email{fsm.fisica@gmail.com}
\affiliation{Universidade Federal do Rio Grande do Norte, Departamento de F\'{i}sica Te\'{o}rica e Experimental, 59300-000, Natal - RN, Brazil.}

\author{Rodrigo F. L. Holanda}
\email{holandarfl@gmail.com}
\affiliation{Universidade Federal do Rio Grande do Norte, Departamento de F\'{i}sica Te\'{o}rica e Experimental, 59300-000, Natal - RN, Brazil.}

\author{Javier E. Gonzalez} \email{javiergonzalezs@academico.ufs.br}
\affiliation{Departamento de F\'{i}sica, Universidade Federal de Sergipe, São Cristóvão, SE 49100-000, Brazil}

\author{L. R. Cola\c{c}o}
\email{colacolrc@gmail.com}
\affiliation{Universidade Federal do Rio Grande do Norte, Departamento de F\'{i}sica Te\'{o}rica e Experimental, 59300-000, Natal - RN, Brazil.}

\author{Rafael C. Nunes}
\email{rafadcnunes@gmail.com}
\affiliation{Instituto de F\'{i}sica, Universidade Federal do Rio Grande do Sul, 91501-970 Porto Alegre RS, Brazil}
\affiliation{Divisão de Astrofísica, Instituto Nacional de Pesquisas Espaciais, Avenida dos Astronautas 1758, São José dos Campos, 12227-010, São Paulo, Brazil}

\begin{abstract}
\noindent 

Testing possible variations in fundamental constants of nature is a crucial endeavor in observational cosmology. This paper investigates potential cosmological variations in the fine structure constant ($\alpha$) through a non-parametric approach, using galaxy cluster observations as the primary cosmological probe. We employ two methodologies based on galaxy cluster gas mass fraction measurements derived from X-ray and Sunyaev-Zeldovich observations, along with luminosity distances from type Ia supernovae. We also explore how different values of the Hubble constant ($H_0$) impact the variation of $\alpha$ across cosmic history. When using the Planck satellite's $H_0$ observations, a constant $\alpha$ is ruled out at approximately the 3$\sigma$ confidence level for $z \lesssim 0.5$. Conversely, employing local estimates of $H_0$ restores agreement with a constant $\alpha$.

\end{abstract}



\maketitle

\section{Introduction}

Searching for a variation of the fundamental constants of nature is also an essential part of our quest to go beyond the $\Lambda$CDM model \cite{uzan2003fundamental,Uzan:2010pm,2017RPPh...80l6902M}. However, the $\Lambda$CDM model has recently encountered some internal observational inconsistencies, specifically the tensions regarding the $H_0$ and $S_8$ parameters \cite{perivolaropoulos2022challenges, Abdalla_2022,di2021realm,kamionkowski2022hubble}. Additionally, it still lacks a satisfactory explanation for longstanding theoretical issues at the intersection of cosmology and particle physics \cite{RevModPhys.61.1,Padmanabhan_2003}

Scenarios where fundamental constants change over cosmic time were first discussed by P. Dirac in his `large numbers hypothesis', which introduced the possibility of a time variation in the gravitational coupling $G$ \cite{dirac1937cosmological}. Over subsequent decades, various theoretical frameworks have been proposed in which fundamental constants can vary in space, time, or both. These include scalar-tensor theories \cite{fujii2003scalar, Naruko:2015zze, hees2014, vandeBruck2015rma}, modified teleparallel gravity \cite{Nunes2016plz, LeviSaid2020mbb}, running vacuum models \cite{Nunes2017}, Bekenstein-Sandvik-Barrow-Magueijo theory \cite{Bekenstein2020wgp, 2012PhRvD85b3514B}, extra dimensions \cite{1997PhR283303O}, and dynamical and interacting dark energy models \cite{2016PhRvD93b3506M, 2015PhRvD91j3501M, 2017PhRvC96d5802M, Liu2021mfk, vonMarttens:2018iav, vonMarttens:2018bvz}, among others. Therefore, testing the stability of fundamental constants on local and cosmological scales offers a powerful means of probing fundamental physics \cite{martins2002cosmology, 2017RPPh...80l6902M}.

The increase and improvement of astronomical data have made this task possible \cite{2020JCAP...06..036G, liu2023, gupta2021, isaac2021, 2016JCAP...08..055H, 2022JCAP...08..062C, ade2015planck, lee2021, 2024arXiv240403123J, lemos2024search, kalita2024constraining, baryakhtar2024varying, jiang2024constraints, jiang2024DESI}. In particular, due to its dimensionless nature, potential space-time variations in the value of the fine structure constant ($\alpha \equiv \frac{e^2}{\hbar c}$, where $e$ is the unit electron charge, $\hbar = h/2\pi$ is the reduced Planck constant, and $c$ is the speed of light) have been investigated using various astrophysical and cosmological probes. These include the Cosmic Microwave Background (CMB) \cite{ade2015planck, Hart:2017ndk, Hart:2019dxi, Hart:2021kad, Hart2022agu, Lynch2024gmp, Wang2022wug}, white dwarf observations \cite{2020IAUS..357...45L, 2017Univ....3...32B}, spectral observations of very high redshift quasars \cite{daFonseca:2022qdf, Webb:2010hc, Murphy:2017xaz, Wilczynska:2020rxx, Milakovic:2020tvq, Murphy:2021xhb}, Big Bang Nucleosynthesis (BBN) \cite{2020A&A...633L..11C, Deal:2021kjs}, strong gravitational lensing systems \cite{2021EPJC...81..822C, li2018testing, 2023GReGr..55..124L}, helium abundance \cite{2023PhRvD.108b3525S, Seto:2023yal}, galaxy clusters \cite{Galli_2013, 2016JCAP...08..055H}, extreme gravitational environments around black holes \cite{Hees:2020gda, Antoniou:2017acq, Silva:2017uqg}, among others. Most results are consistent with no significant variation in $\alpha$. Furthermore, local methods have also been employed to obtain stringent bounds on potential variations in $\alpha$, such as studies of the Oklo natural nuclear reactor \cite{Damour1996zw, Petrov2005pu, Gould2006qxs, Onegin2010kq} and laboratory measurements of atomic clocks with different atomic numbers \cite{2004PhRvL92w0802F, PhysRevLett93170801, 101126science1154622}. In addition to its significant impact on fundamental physics, this possibility has recently been suggested as a potential solution to some cosmological challenges, such as the Hubble tension \cite{Hart:2021kad, franchino2021cosmological, zhang2023mirror,lee2023takes, chluba2023varying, vacher2024incompatibility}.

On the other hand, the authors of Refs.\cite{hees2014,Minazzoli:2014xua} demonstrated that variations in the fine structure constant, violations of the cosmic distance duality relation (CDDR), and deviations from the standard evolution of the CMB temperature law are intimately linked to modifications of gravity due to the presence of an extra scalar field ($\phi$) non-minimally coupled to usual matter fields $\Psi_i$. Such a coupling affects the entire electromagnetic sector, and is studied through the action:

\begin{equation}
S_{m} = \sum_i \int d^4x \sqrt{-g} h_i(\phi) \mathcal{L}_i(g_{\mu\nu}, \Psi_i)\,,
\end{equation}
where $\mathcal{L}_i$ are the Lagrangians for different matter fields $\Psi_i$, and $h_i(\phi)$ represents the non-minimal couplings between $\phi$ and $\Psi_i$. These changes are tied to the time evolution of $h(\phi(t))$, and when $h_i(\phi) = 1$, general relativity is recovered. Also note that this nonminimal coupling is largely motivated by multiple alternative theories of gravity. For instance, the low-energy action of String Theories \cite{Damour1994zq, damour1994string, gasperini2001, minazzoli2014}, the generalized Cosmological Chameleons \cite{Brax:2004qh, Brax:2007ak, Weltman:2008fp, Ahlers:2007st}, theories of Kaluza-Klein with additional compactified dimensions \cite{fujii2003scalar, overduin1997}, in the context of axions \cite{PhysRevD.16.1791,DINE1981199,KAPLAN1985215}, the varying-$\alpha$ Bekensteiin-Sandvik-Barrow-Magueijo Theory \cite{Bekenstein2020wgp, Sandvik2001rv, Barrow2011kr, Barrow2013uza}, in and others \cite{Harko2012hm,Minazzoli2013ara}.

A specific case within this context is the Runaway Dilaton model \cite{2019PhRvD.100l3514M, 2015PhLB..743..377M, hees2014, Martinelli2015cza, Damour1994zq, Damour1992kf, Damour2002nv}, inspired by string theory and capable of describing time variations of $\alpha$ near the present day. The coupling of the dilaton field to hadronic matter is the key parameter in studying $\alpha$ variation. Recent observational limits on $\alpha$ variations have been explored using different methodologies. Notable studies include Ref.\cite{2017PhRvD95h4006H}, which used X-ray gas mass fraction measurements combined with SNe Ia observations, and Ref.\cite{2016JCAP05047H}, which considered galaxy cluster (GC) gas mass fraction measurements obtained through X-ray and Sunyaev-Zeldovich effect (SZE) observations. These works established a direct relationship between galaxy cluster observations, possible variations of the fine structure constant, and violations of the CDDR.

In this paper, we allow for a possible time variation in $\alpha$, such that $\alpha(z) = \alpha_0 \phi(z)$, using a non-parametric method to reconstruct the $\phi(z)$ function considering galaxy cluster observations as the primary cosmological tool. Additionally, as a complementary probe, we utilize $D_L(z)$ measurements from SNe Ia observations drawn from the Pantheon sample. 
We employ two distinct methodologies based on those demonstrated in Refs.\cite{2016JCAP...08..055H,2017PhRvD95h4006H} (see next section). For the first methodology, we utilize a combination of $f_{\text{gas}}$ from X-ray measurements and $D_L(z)$ from SNe Ia. For the second method, we combine $f_{\text{gas}}^{\text{X-ray}}$ and $f_{\text{gas}}^{\text{SZE}}$ to explore the behavior of $\phi(z)$. It is worth highlighting that we do not consider the runaway dilaton model or any specific model describing $\alpha$ evolution. We also explore the influence of different values of the Hubble constant on the $\alpha(z)$ constraints. 

This paper is organized as follows: In Section \ref{Methodology}, we describe the methodologies we  employ for our purposes. The data are detailed in Section \ref{samples}. The analyses and the discussions are presented in Section \ref{results}. Finally, we conclude the paper in Section \ref{final}.

\section{Methodology}
\label{Methodology}

Particularly, two types of geometric distances are of great importance for observational cosmology: the luminosity distance, $D_L$, and the angular diameter distance, $D_A$. The former measures the distance of an object based on the decrease in its brightness with distance, while the latter relates to the angular size of the object projected on the celestial sphere. Both distances depend on the redshift $z$ and are related to each other by the so-called CDDR \cite{CDDR,Bassett:2003vu}:

\begin{equation}
    \frac{D_L (z)}{(1+z)^2 D_A (z)}=1.
\end{equation}
This equation is a version of Ethertington's reciprocity law in the context of astronomical observation. It is trivially satisfied in a Friedmann-Lemaître-Robertson-Walker (FLRW) framework, nevertheless, it is completely general and valid for all cosmological models based on Riemannian geometry \footnote{Such generality makes the CDDR as being of great importance in observational cosmology, and any deviation might indicate new physics or the presence of systematic errors in observations \cite{Ellis2007}.}. It only requires the observer and the source to be connected by null geodesics, and the number of photons to be conserved over the cosmic evolution.

As usual, the CDDR can be parameterized to check its validity through astronomical data as
\cite{2016JCAP...08..055H,holanda2016probing,holanda2017probing,mukherjee2021assessment}:

\begin{equation} \label{cddr_violation}
    \frac{D_L(z)}{D_A(z) (1 + z)^2} \equiv \eta(z).
\end{equation}
However, in Ref.\cite{hees2014}, the authors demonstrate that within the framework of non-minimal multiplicative couplings between an extra scalar field and matter fields, variations of the fine-structure constant and violations of the CDDR are linked by

\begin{equation} \label{delta_alpha}
    \frac{\Delta \alpha(z)}{\alpha_0} = \frac{\alpha(z) - \alpha_0}{\alpha_0} = \eta^2(z) - 1.
\end{equation}
This equation is essential to this work since we investigate variations in $\alpha$ considering $\alpha = \alpha_0 \phi(z)$, and some observables depend on both the $\phi(z)$ and $\eta(z)$ functions. We outline the methods and samples in the following sections.

\subsection{Method I: Using $f_{\text{gas}}$ from X-ray observations and SNe Ia} 

This subsection discusses the methodology presented in Ref.\cite{2017PhRvD95h4006H}. The gas mass fraction is defined as \cite{sasaki1996,2011ARAA..49..409A,Mantz_2021}

\begin{equation}
    f_{\text{gas}} = \frac{M_{\text{gas}}}{M_{\text{tot}}},
\end{equation}
where $M_{\text{tot}}$ represents the total mass and $M_{\text{gas}}$ denotes the gas mass obtained by integrating the gas density model, expressed in spherical coordinates as:

\begin{equation}
    M_{\text{gas}}(< V) = 4 \pi \int_0^R \rho_{\text{gas}} r^2 dr.
\end{equation}

The intracluster gas originates from primordial gas composed of hydrogen ($H$) and helium ($He$). Thus,

\begin{equation}
    n_H = \left( \frac{2 X}{1 + X} \right) n_e(r)
\end{equation}
and
\begin{equation}
    n_{He} = \left[ \frac{1 - X}{2(1+X)} \right] n_e(r),
\end{equation}
where $X$ denotes the hydrogen abundance, and $\rho_{\text{gas}} = \rho_H + \rho_{He}$. Therefore,

\begin{equation}
    \rho_{\text{gas}} = \frac{2 n_{e0} m_H}{(1 + X)} \left( 1 + \frac{r^2}{r_c^2} \right)^{- 3 \beta / 2},
\end{equation}
where $m_H$ represents the hydrogen mass. We use the isothermal spherical $\beta$-model to describe the electron density \cite{cavaliere1978}:

\begin{equation} \label{profile}
    n_e(r) = n_{e0} \left[ 1 + \left( \frac{r}{r_c} \right)^2 \right]^{-3 \beta / 2}.
\end{equation}
Thus, it leads to:

\begin{equation} \label{m_gas}
    M_{\text{gas}}(< V) = \frac{8 \pi n_{e0} m_H r_c^3}{(1 + X)} I_M(R/r_c, \beta),
\end{equation}
where

\begin{equation}
    I_M(R/r_c, \beta) \equiv \int_0^{R/r_c} (1 + x^2)^{-3 \beta / 2} x^2 dx 
\end{equation}
with $x = r/r_c$. Assuming hydrostatic equilibrium and isothermality, and using Eq. \eqref{profile}, the total mass can be expressed as:

\begin{equation} \label{m_tot}
    M_{\text{tot}}(< R) = \frac{3 \beta k_B T_e}{\mu G m_H} \left[ \frac{R^3}{(r_c^2 + R^2)} \right],
\end{equation}
where $T_e$ is the temperature of the intracluster medium obtained from the X-ray spectrum, $\mu$ is the mean molecular weight, and $m_p$ is the proton mass. Finally, combining Eq. \eqref{m_gas} and Eq. \eqref{m_tot}, we obtain:

\begin{equation}
    f_{\text{gas}} = \frac{8 \pi m_H^2 \mu G n_{e0}}{3 (1 + X) \beta k_B T_e} \left[ \frac{r_c^5 + r_c^3 R^2}{R^3} \right] I_M(R/r_c, \beta).
\end{equation}

The parameter $n_{e0}$ can be determined through observations such as the Sunyaev-Zel'dovich effect and X-ray surface brightness. In this methodology, we focus on X-ray observations and explicitly detail the dependence of $f_{\text{gas}}^{\text{X-ray}}$ on $\alpha$.

\subsubsection{X-ray observations}

The X-ray emission from galaxy clusters is mainly from the Intra Cluster Medium (ICM) gas through thermal Bremsstrahlung emission \cite{rybicki1991, sarazin1986}, where the bolometric luminosity is given by 

\begin{equation}
    L_x = 4 \pi \int_0^R \frac{d L_x}{dV} r^2 dr,
\end{equation}
where $d L_x/dV$ is given by

\begin{equation} \label{emissivity}
    \frac{dL_x}{dV} = \alpha^3 \left( \frac{2 \pi k_B T_e}{3 m_e} \right)^{1/2} \frac{2^4 \hbar^2}{3 m_e} n_e \left( \sum_i Z_i^2 n_i g_{B_i} \right).
\end{equation}

We consider the isothermal spherical $\beta$-model and, as mentioned before, the intracluster medium is composed of hydrogen and helium. Thus, one obtains:

\begin{eqnarray} \label{LX}
    L_x &=& \left( \frac{2 \pi k_B T_G}{3 m_e}  \right)^{1/2} \frac{2^4 e^6}{3 \hbar m_e c^3} g_B(T_G) \frac{2}{(1 + X)} 4 \pi n_{e0} \nonumber \\
    &\times & \int_{0}^{R} \left( 1+ \frac{r^2}{r_{c}^{2}}  \right)^{-3\beta}r^2 dr.
\end{eqnarray}
Defining 

\begin{equation}
    I_L(R/r_c, \beta) \equiv \int_0^{R/r_c} (1 + x^2)^{-3 \beta} x^2 dx,
\end{equation}
where $x = r/r_c$, it is possible to rewrite the Eq. \eqref{LX} taking into consideration $\alpha$ and obtain \cite{2017PhRvD95h4006H}: 

\begin{eqnarray}
 L_x &=& \alpha^3   \left( \frac{2 \pi k_B T_G}{3 m_e}  \right)^{1/2}  \frac{2^4 \hbar^2}{3 m_e} g_B(T_G) \frac{2}{(1  + X)} \nonumber \\ 
 &\times & 4 \pi n_{e0}^2 D_{A}^{2} \theta_c^2 r_c  I_L(R/r_c, \beta).
\end{eqnarray}

The quantity $L_x$, the total X-ray energy per second emitted from the galaxy cluster, is not directly observable. Actually, the quantity observable is the X-ray flux, which is given by \cite{sasaki1996, 2011ARAA..49..409A,  Mantz_2021}:

\begin{equation}
    F_x = \frac{L_x}{4 \pi D_L^2}.
\end{equation}
As one may see, $n_{e0} \propto \alpha^{-3/2} D_L / D_A$. Therefore, if $\alpha/\alpha_0 = \phi (z)$ and the CDDR varies as $\eta(z)$, the gas mass fraction measurements derived from X-ray data are affected by deviations of $\alpha$ and $\eta$ as \cite{2016JCAP05047H}:

\begin{equation} \label{f_xray_alpha}
    f^{\text{th}}_{\text{X-ray}} \propto [\phi(z)]^{-3/2} \eta(z).
\end{equation}

As a cosmological tool, the X-ray gas mass fraction is obtained through the following expression \cite{Allen2008,allen2011,corasaniti2021}:

\begin{equation} \label{gas_mass_fraction}
    f_{\text{gas}} = A(z) K \gamma \frac{\Omega_b}{\Omega_m} \left[ \frac{D_L^*}{D_L} \right]^{3/2},
\end{equation}
where  $A(z)$ represents the angular correction factor, which is close to unity for all cosmologies and redshifts of interest \cite{Mantz2014xba}, $\Omega_m$ is the total matter density parameter and $\Omega_b$ is the baryon density parameter. The gas mass fraction measurements performed by the Ref.\cite{Mantz2014xba} depend on the luminosity distance to each galaxy cluster in the sample. In general, a fiducial cosmological model ($D^*_L$) is used for this purpose (typically the flat $\Lambda$CDM model with Hubble constant $H_0 = 70$ km s$^{-1}$ Mpc$^{-1}$ and $\Omega_m=0.3$). Thus, the ratio inside the brackets takes it into account, making the analyses with gas mass fraction measurements independent of the fiducial cosmological model. { The factor $\gamma  = \gamma_0$ $( M_{2500}/3\times 10^{14} M_{\odot} )^{a}$, where we assumed here $a = 0.025 \pm 0.0353$} and $\gamma_0 = 0.79 \pm 0.07$ \cite{mantz2022} is the depletion factor; $K(z)$ represents the calibration constant, which quantifies inaccuracies in instrument calibration and bias in the masses measured due to substructures, bulk motions and non-thermal pressure in the cluster gas \cite{Mantz2014xba,Allen2008}; it is assumed here to be $K = 0.93 \pm 0.11$ \cite{mantz2022}. In the following sections, we assume $\Omega_b / \Omega_m$ from galaxy clustering \cite{2024arXiv240319236K}.

Therefore, employing the relation between a varying-$\alpha$ and a departure of the CDDR \eqref{delta_alpha}, along with the relation \eqref{f_xray_alpha}, we may derive:

\begin{equation} \label{phi_2}
    \phi(z) = \frac{A(z) K \gamma \frac{\Omega_b}{\Omega_m} [ D_L^* / D_L]^{3/2}}{f_{\text{gas}}^{\text{X-ray}}}.
\end{equation}
This key equation, eq. (\ref{phi_2}), will be used to reconstruct $\phi(z)$ based on $f_{\text{gas}}$ measurements and SNe Ia luminosity distances.

\subsection{Method II: Using $f_{\text{X-ray}}$ and $f_{\text{SZE}}$}

In this subsection, we discuss the methodology presented in Ref.\cite{2016JCAP05047H}. Allowing $\alpha$ to vary together with the CDDR by \eqref{delta_alpha}, we establish a specific connection between $f_{\text{X-ray}}$ and $f_{\text{SZE}}$ following \cite{2016JCAP05047H}. According to the authors, if $\phi(z) \neq 1$ and $\eta(z) \neq 1$, we derive $f^{\text{th}}_{\text{SZE}} = \phi(z)^{-2} f_{\text{SZE}}$ and $f^{\text{th}}_{\text{X-ray}} = \phi(z)^{-3/2} \eta(z) f_{\text{X-ray}}$, representing the modified gas mass fraction equations obtained from the Sunyaev-Zel'dovich effect and X-ray observations, respectively\footnote{The index th indicates the theoretical quantity}.

Since $f_{\text{SZE}}$ and $f_{\text{X-ray}}$ aim to measure the same physical quantity, a consistent agreement between them is expected to be as $f^{\text{th}}_{X-ray} / f^{\text{th}}_{\text{SZE}} = 1$. Therefore, the expression relating to X-ray and SZE observations is:

\begin{equation} \label{fsze_xray}
    f_{\text{SZE}} = \phi(z)^{1/2} \eta(z) f_{\text{X-ray}}.
\end{equation}
Considering Eq. \eqref{delta_alpha}, we obtain:

\begin{equation} \label{alpha_fsze}
    f_{\text{SZE}} = \phi(z) f_{\text{X-ray}}.
\end{equation}
This equation directly examines $\phi(z)$, encompassing influences from potential variations of $\alpha$ and deviations of the CDDR.

\begin{figure*}[htbp]
    \centering
    \begin{subfigure}[b]{0.45\textwidth}
        \centering
        \includegraphics[width=\textwidth]{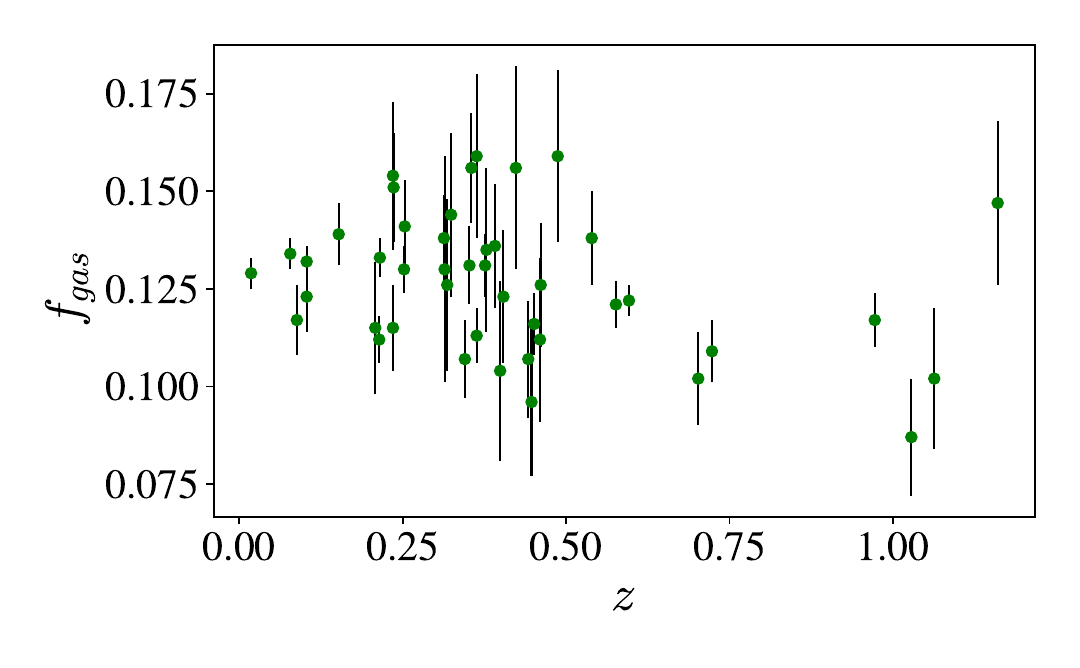}
    \end{subfigure}
    \hfill
    \begin{subfigure}[b]{0.45\textwidth}
        \centering
       \includegraphics[width=\textwidth]{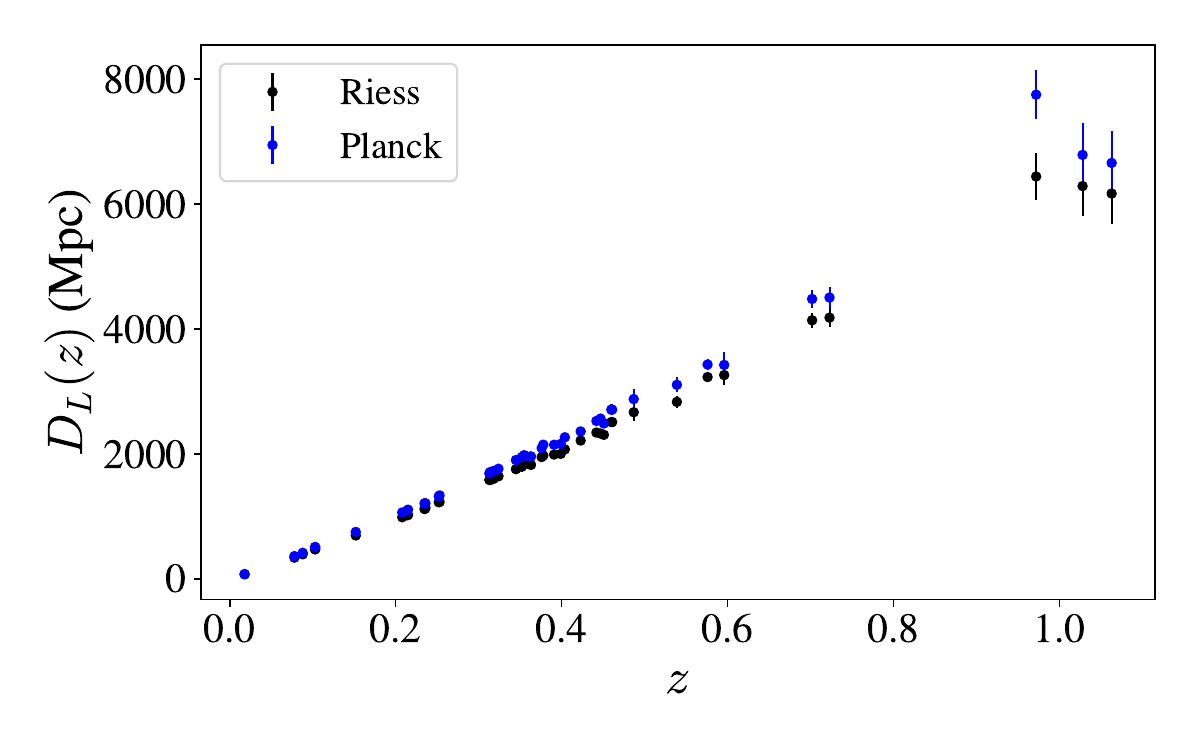}
 \end{subfigure}
 \caption{Left panel: The sample of gas mass fractions adopted in this work.
Right panel: Weighted average luminosity distance measurements derived for our sample and obtained from the Pantheon Sample using $H_0$ values from the Planck and SH0ES collaborations.
}
 \label{fig:fgas+DL}  
\end{figure*}
 
\section{Samples} \label{samples}

Below, we describe the datasets that will be applied in the methodology presented in the previous section:

\begin{itemize}

\item{\bf Gas mass fraction:} We utilize the new sample of $f_{\text{gas}}$ compiled by \cite{mantz2022}, which includes 44 massive, hot, and morphologically relaxed galaxy clusters in the redshift range $0.018 \leq z \leq 1.160$, observed by Chandra (see Fig. \ref{fig:fgas+DL}). The selection of relaxed systems aims to reduce systematic uncertainties and scatter arising from deviations in hydrostatic equilibrium, spherical symmetry, and the need for the intracluster medium to maintain a temperature of at least $5$ keV within the isothermal region of the temperature profile. This approach minimizes systematic uncertainties associated with cluster formation and feedback from active galactic nuclei. Furthermore, the innermost cluster regions were deliberately excluded, allowing $f_{\text{gas}}$ to be calculated in the spherical shell $0.8 \leq r / r_{2500} \leq 1.2$, where $r_{2500}$ denotes the radius at which the mean enclosed mass density equals $2500 \rho_{crit}$.

\item{\bf $f_{\text{SZE}}$ and $f_{\text{X-ray}}$:} For the gas mass fraction from SZE and X-ray observations, we utilize a dataset compiled by \cite{laroque2006x}. This dataset consists of 38 gas mass fractions of massive galaxy clusters within the redshift range $0.14 \leq z \leq 0.89$, derived from Chandra X-ray data and OVRO/BIMA interferometric SZE measurements. In their work, \cite{laroque2006x} employed three gas distribution models: an isothermal $\beta$-model fit to the X-ray data beyond $100$ kpc combined with all SZE data, a non-isothermal double-$\beta$ model fit to both X-ray and SZE datasets, and an isothermal $\beta$-model fit solely to the SZE spatial data. Their analyses demonstrated that the core could be adequately addressed either by excluding it in fits of the X-ray data (the $100$ kpc-cut model) or by employing a non-isothermal double-$\beta$ model to characterize the intracluster gas. In our analysis, we adopt the non-isothermal double-$\beta$ model for the gas distribution (see Table 5 in \cite{laroque2006x}), and the 3D temperature profile was modeled assuming the intracluster medium (ICM) is in hydrostatic equilibrium with an NFW dark matter density distribution.

However, several clusters in this study showed a questionable reduced $\chi^2$ when described by the hydrostatic equilibrium model. These clusters include Abell 665, ZW 3146, RX J1347.5-1145, MS 1358.4+6245, Abell 1835, MACS J1423+2404, Abell 1914, Abell 2163, and Abell 2204. Therefore, we excluded these clusters from our final dataset, resulting in a subsample of 29 clusters (see Fig. \ref{fig:fxray_fsze}).

\begin{figure}[htbp]
\includegraphics[width=0.5\textwidth]{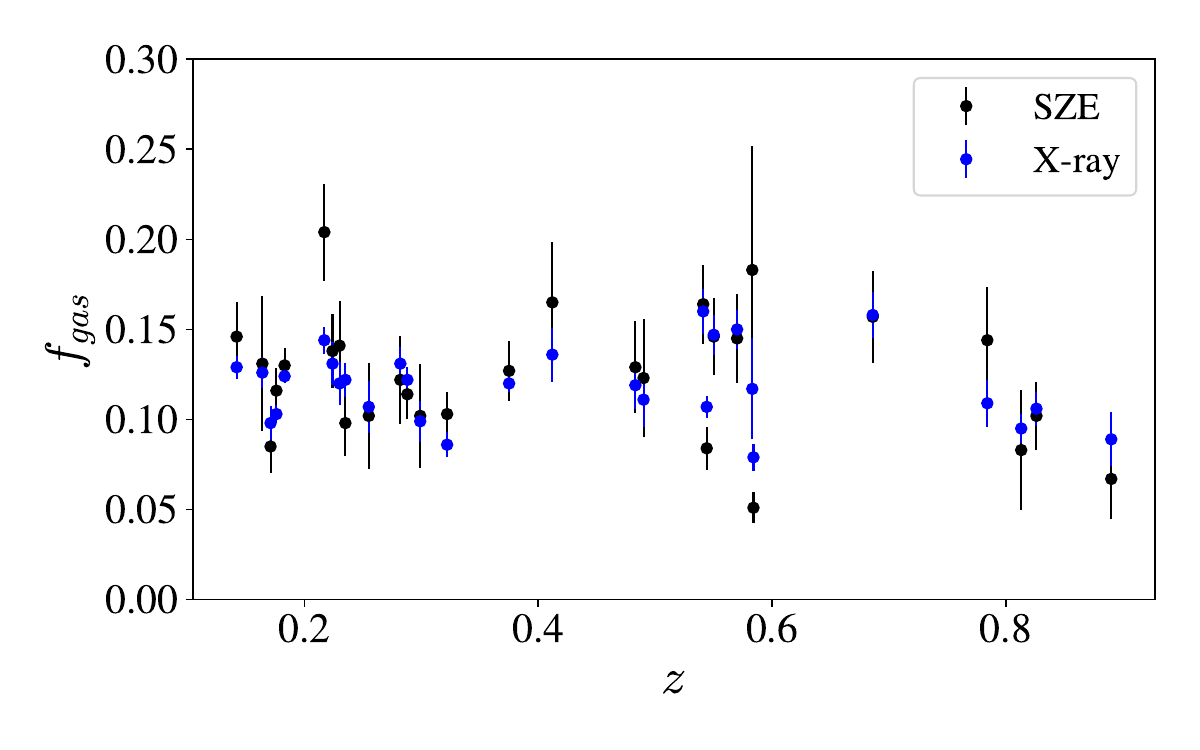}
\caption{Gas mass fraction measurements from Sunyaev-Zel'dovich effect and X-ray brightness used in Method II.}
\label{fig:fxray_fsze}
\end{figure}

\item{\bf Luminosity Distances from SNe Ia:} We construct two datasets of $D_L(z)$ from the Pantheon sample \cite{Scolnic:2017caz}, which consists of spectroscopically confirmed type Ia supernovae within a redshift range $0.01 \leq z \leq 2.3$. The luminosity distance for each supernova is obtained using the relation

\begin{equation}
D_L = 10^{(\bar{m}_b - M_b - 25)/5} \ \text{Mpc},
\end{equation}
where $\bar{m}_b$ and $M_b$ are the apparent and absolute magnitudes, respectively. In this study, we create two datasets of SNe Ia luminosity distances. The first dataset assumes $H_0 = 67.4 \pm 0.5$ km/s/Mpc based on Planck-CMB estimates \cite{Planck:2018vyg}. The second dataset assumes $H_0 = 73.04 \pm 1.04$ km/s/Mpc based on SH0ES estimates \cite{riess2022comprehensive}.

The Pantheon data includes the covariance matrix of $\bar{m}_b$ for each supernova, which needs to be transformed into a covariance matrix for $D_L$ using the relation:

\begin{eqnarray}
\text{cov}({\bf D_L, D_L}) = \left(  \frac{\partial {\bf D_L}}{\partial {\bf m_b}} \right) \text{cov}({\bf m_b, m_b}) \nonumber \\
         \times \left(  \frac{\partial {\bf D_L}}{\partial {\bf m_b}} \right)^T,
     \end{eqnarray}
where $\left(  \frac{\partial {\bf D_L}}{\partial {\bf m_b}} \right)$ is the Jacobian matrix of the transformation and the variables in bold correspond to vector representations of each data set.
     
To perform our analysis, we match SNe Ia and galaxy clusters in identical redshift bins. For each galaxy cluster, we select SNe Ia with redshifts satisfying $\mid z_{GC} - z_{SN} \mid \leq 0.005$. We then calculate the following weighted average for the SNe Ia data, considering the covariance matrix to reduce uncertainties and account for correlations between data points:

\begin{equation}
\label{DL_mean}
\bar{D}_L = \frac{\sum_{i,j} (D_{L_i} w_{ij})}{\sum_{i,j} w_{ij}},
\end{equation}
where $D_{L_i}$ are the individual luminosity distances of selected SNe Ia, and $w_{ij} = (\text{cov}^{-1}(\mathbf{D_L, D_L}) )_{ij}$ represents elements of the inverse covariance matrix.

This process results in 43 measurements of $\bar{D}_L$ that correspond to the galaxy cluster sample \cite{mantz2022} (see Fig. \ref{fig:fgas+DL}). We can observe that for high values of $z$, there is a significant difference in the estimates of $\bar{D}_L$. This difference arises from the rescaling of these estimates using different values of $H_0$, which are currently well known to be in tension by more than 5$\sigma$ confidence level (CL).

The weighted average luminosity distances are correlated due to the inherent correlations in the supernova data and the overlap of supernovae used to calculate different $\bar{D}_L$ values. It can be shown that the covariance matrix of the averaged luminosity distances satisfies the following relation \cite{colacco2022varying}:

\begin{equation*}
\text{cov}(\bar{D}_{L}^i, \bar{D}_{L}^j) =
\end{equation*}
\begin{equation}
\sum_{\alpha,\gamma}^{n^i, n^j} \frac{\text{cov}(D^i_{L_\alpha}, D^j_{L_\gamma}) \left[ \sum_{\beta}^{n^i} w_{\alpha \beta}^i \right] \left[ \sum_{\sigma}^{n^j} w_{\gamma \sigma}^j \right]}{\left[ \sum_{\sigma} ^{n^i} \sum_{\beta}^{n^i} w_{\sigma \beta}^i \right] \left[ \sum_{\sigma} ^{n^j} \sum_{\beta}^{n^j} w_{\sigma \beta} ^j\right]},
\end{equation}
where the superindexes represent the sets of SNe considered to calculate the luminosity distances in Eq. \eqref{DL_mean}, whereas the subindexes correspond to the i-th term in each set and $n$ is the number of data points.

\end{itemize}

\begin{figure*}[htb]
    \centering
    \begin{subfigure}[t]{0.45\textwidth}
        \centering
        \includegraphics[width=\textwidth]{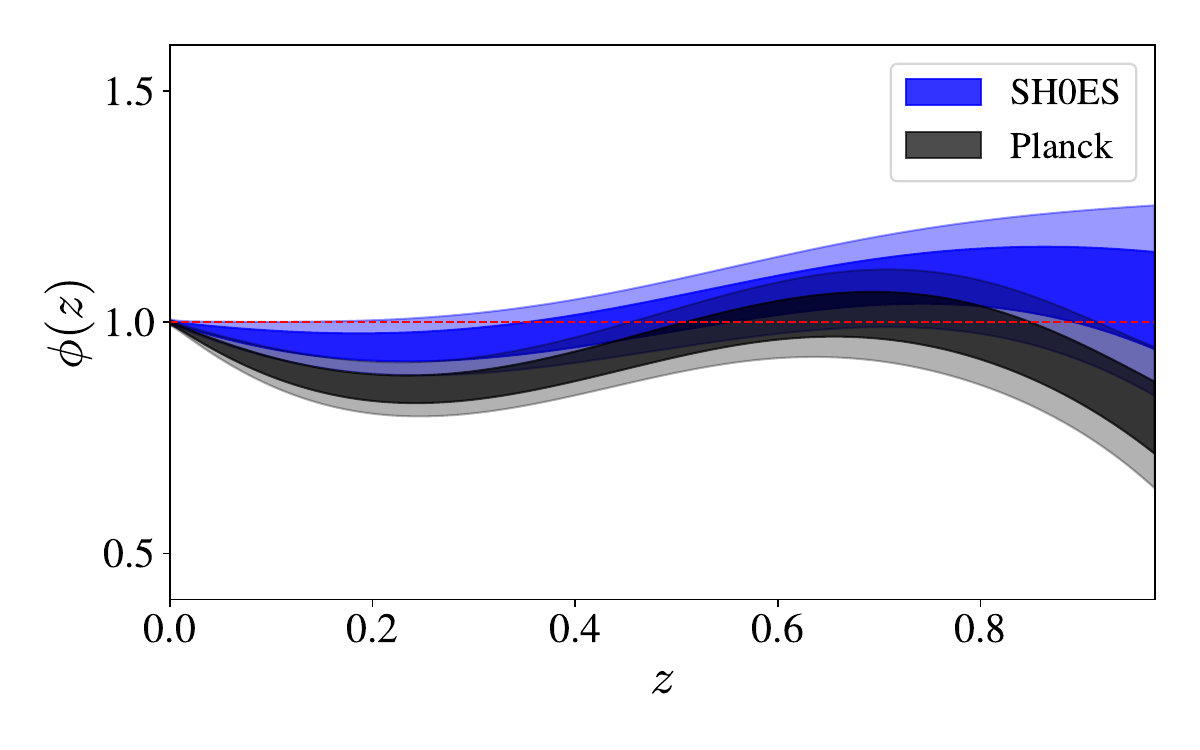}
        \label{fig:figura_fgas_mantz}
    \end{subfigure}
    \hfill
    \begin{subfigure}[t]{0.45\textwidth}
        \centering
        \includegraphics[width=\textwidth]{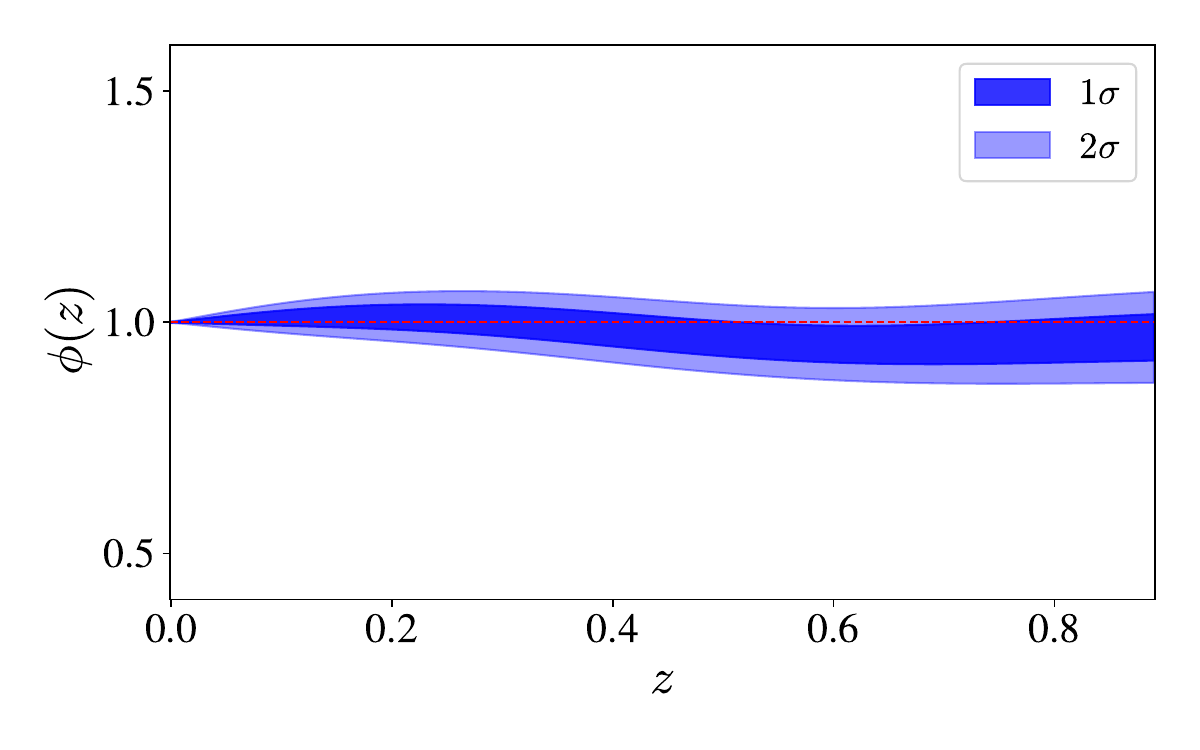}
        \label{fig:figura_fsze}
    \end{subfigure}

    \vspace{0.5cm} 
    
    \caption{Left panel: Reconstruction of $\phi(z)$ for Method I. Right panel: Reconstruction of $\phi(z)$ for Method II, which utilizes $f_{\text{X-ray}}$ and $f_{\text{SZE}}$ measurements. The dashed line represents the scenario with no variation in $\alpha$.
}
    \label{RESULT_}
\end{figure*}

\begin{table*}[htb!]
\resizebox{\columnwidth}{!}{%
\setcounter{table}{0} 
\begin{tabular}{l c c c}
\hline
\textbf{Method} & $\alpha(z = 0.3)\times 10^{-3}$ & $\alpha(z = 0.6)\times 10^{-3}$ & $\alpha(z = 0.9)\times 10^{-3}$ \\ \hline
Method I ($H_0$ from Planck) & $(6.295 \pm 0.219)$ & $(7.326 \pm 0.299)$ & $(6.477 \pm 0.474)$ \\
Method I ($H_0$ from SH0ES) & $(6.961 \pm 0.241)$ & $(7.722 \pm 0.307)$ & $(7.871 \pm 0.606)$ \\
Method II & $(7.311 \pm 0.241)$ & $(6.961 \pm 0.285)$ & $(7.066 \pm 0.365)$ \\ \hline
\end{tabular} 
}
\caption{Predicted values for $\alpha(z)$ at three different $z$ values for each case of the analysis at a $68\%$ CL. { For reference, the current value of the fine structure constant ($z=0$) is $\approx 7.297 \times 10^{-3}$.} 
} 
\label{tabela}
\end{table*} 

\section{Analysis and Main Results}
\label{results}

To reconstruct $\phi(z)$ using a Gaussian Process (GP), we transform the observational data sets described above into mean values of the function $\phi(z)$ and its associated covariance matrix. Each data point is assumed to follow a Gaussian distribution, and this transformation is carried out using Monte Carlo sampling\footnote{Due to the high errors in the gas mass fraction data, the Gaussian propagation of errors using partial derivatives is not adequate.}. For SNe Ia, where data are correlated, we consider a joint multivariate Gaussian distribution; for other quantities, we assume one-dimensional normal distributions. A GP provides a method to reconstruct a function from observational data without assuming a specific model. At each point $z$, the reconstructed function $f(z)$ is characterized by a normal distribution. The values of the function at different points are interconnected through a covariance function $k(x,x')$. The GP is performed by choosing a prior mean function and a covariance function (see \cite{Seikel:2012uu} for more details). For all reconstructions, we chose zero as the prior mean function to avoid potentially biased results and used a standard Gaussian kernel as the covariance function between two data points separated by a redshift distance $z-z'$:


\begin{equation} \label{kernel}
    k(z,z') = \sigma^2 \exp \left( -  \frac{(z - z')^2}{2l^2} \right),
\end{equation}
where the parameter $l$ corresponds to the scale in $z$ over which significant changes occur in the function values, while $\sigma$ denotes the typical magnitude of these changes. We also require that $\alpha(z = 0) = \alpha_0$. The hyperparameters are optimized by maximizing the loglikelihood: 

\begin{eqnarray}
\ln {\mathcal{L}}=-\frac{1}{2} \bm {\bm (\phi)}^T [\bm K(\bm {z}, \bm {z}) + \bm C]^{-1} \bm {\bm (\phi)} \nonumber \\ 
- \frac{1}{2} \ln | \bm K(\bm z, \bm z) + \bm C| - \frac{n}{2} \ln 2 \pi,
\end{eqnarray}
where $\bm \phi$ and $\bm z$ are the vectors of the dependent and independent data variables, respectively, $\bm K(\bm z,\bm z)$ is the covariance matrix used to describe the data as a GP and its elements are calculated with Eq. \eqref{kernel}, $\bm C$ is the covariance matrix of the data which contains the errors of the function, and $n$ is the number of data points. We used the GaPP\footnote{https://github.com/carlosandrepaes/GaPP} package to perform the GP. The analyses for each method are described as follows.

For Method I, we utilize a gas mass fraction sample calculated at $r_{2500}$ along with two samples of SNe Ia for $D_L(z)$, corresponding to $H_0$ values consistent with estimates from the Planck satellite and the SH0ES team. This allows us to obtain $\phi(z)$ and its respective covariance matrix from Eq. \eqref{phi_2} by considering all quantities in a Monte Carlo sampling. We observe that the Gaussianity remains a good approximation for the random variable $\phi$ corresponding to gas mass fraction data with relative errors smaller than 15\%. However, for $f_{\text{gas}}$ data with higher relative errors, the Gaussianity of the obtained variable $\phi$ is broken, and it is not possible to describe those data using a Gaussian Process (GP). Therefore, we consider gas mass fraction data up to 15\% relative error to construct the $\phi$ sample. Using the Gaussian Process method, we reconstruct $\phi(z)$ according to Eq. \eqref{phi_2}. 

The results are shown in the left panel of Fig. \ref{RESULT_}. Firstly, it is important to note a possible minimal oscillatory behavior in the reconstruction due to the large dispersion in the $f_{\text{gas}}$ data. Thus, this behavior is data-driven. For both reconstructions, i.e., with prior $H_0$ from CMB and SH0ES, we note good agreement with $\phi = 1$ for $z > 0.5$. For $z < 0.5$, we noticed a significant drop in the amplitude of $\phi(z)$ in the CMB case, leading to deviations in $\phi(z)$ by more than $2\sigma$ CL for $z \in [0, 0.4]$. In other words, the electromagnetic coupling via $\alpha$ becomes weaker at late times. On the other hand, imposing the SH0ES prior, the amplitude of $\phi(z)$ tries to increase, restoring agreement with $\phi = 1$. Therefore, depending on the value that $H_0$ may have, this will certainly have an impact on the evolution of the scalar field and consequently on the value of $\alpha$.

For Method II, we extract the values of $\phi(z)$ and their corresponding errors from $f_{\text{X-ray}}$ and $f_{\text{SZE}}$ measurements. Using these values, we reconstruct the function according to \eqref{alpha_fsze} using the Gaussian Process. The outcome is depicted in the right panel of Fig. \ref{RESULT_}. The result demonstrates agreement with a constant $\alpha$ at $2 \sigma$ confidence level for every $z$ range analyzed. Table \ref{tabela} displays the values of $\alpha(z)$ obtained from the reconstruction of $\phi(z)$ when evaluated in bins of $\Delta z = 0.3$, within the range used in our reconstructions.

\section{Conclusions}
\label{final}

In this paper, we have investigated the potential time variation of the fine structure constant, $\alpha$, using a non-parametric approach. Our study employed two distinct methodologies, detailed in Section II. Unlike previous approaches that often rely on specific models for $\alpha$ evolution, our method remains model-independent, providing greater flexibility in interpreting observational data. Our findings, as detailed in Section IV, highlight that the estimated values of $\alpha$ can be sensitive to the adopted value of $H_0$. This sensitivity underscores the ongoing debate regarding potential new physics underlying the tension observed in cosmological parameters, which could directly influence the evolution of $\alpha(z)$. In conclusion, our non-parametric analysis supports the prevailing understanding of $\alpha$ as a constant of nature, validated by independent cosmological probes. Future investigations could enhance our analysis with more precise datasets, potentially from upcoming surveys, to further refine our understanding of fundamental constants and their evolution across cosmic epochs.

\begin{acknowledgments}
\noindent RFLH thanks to CNPq support under the project No. 309132/2020-7; LRC also thanks to CNPQ support under the project No. 169625/2023-0; RCN also thanks the financial support from the CNPq under project No. 304306/2022-3, and the FAPERGS for partial financial support under project No. 23/2551-0000848-3.
\end{acknowledgments}

\bibliography{PRD}

\end{document}